\documentclass[journal,english]{IEEEtran}
\makeatletter
\long\def\@makecaption#1#2{\ifx\@captype\@IEEEtablestring%
\footnotesize\begin{center}{\normalfont\footnotesize #1}\\
{\normalfont\footnotesize\scshape #2}\end{center}%
\@IEEEtablecaptionsepspace
\else
\@IEEEfigurecaptionsepspace
\setbox\@tempboxa\hbox{\normalfont\footnotesize {#1.}~~ #2}%
\ifdim \wd\@tempboxa >\hsize%
\setbox\@tempboxa\hbox{\normalfont\footnotesize {#1.}~~ }%
\parbox[t]{\hsize}{\normalfont\footnotesize \noindent\unhbox\@tempboxa#2}%
\else
\hbox to\hsize{\normalfont\footnotesize\hfil\box\@tempboxa\hfil}\fi\fi}
\makeatother

\usepackage[utf8]{inputenc}
\IEEEoverridecommandlockouts
\usepackage[autostyle=true]{csquotes} % Required to generate language-dependent quotes in the bibliography
\usepackage{fancyhdr}
\usepackage{graphicx}
\usepackage[english]{babel}
\usepackage{color}
\usepackage{hyperref}
\usepackage{wrapfig}
\usepackage{array}
\usepackage{multirow}
\usepackage{adjustbox}
\usepackage{nccmath}
\usepackage{subfigure}
\usepackage{amsfonts,latexsym} % para tener disponibilidad de diversos simbolos
\usepackage{enumerate}
\usepackage{booktabs}
\usepackage{float}
\usepackage{threeparttable}
\usepackage{array,colortbl}
\usepackage{ifpdf}
\usepackage{rotating}
\usepackage{cite}
\usepackage{stfloats}
\usepackage{url}
\usepackage{listings}
\let\vec\mathbf
\usepackage{euscript}
\usepackage{mathdots}
\usepackage{listings}
\usepackage{xcolor}

\definecolor{codegreen}{rgb}{0,0.6,0}
\definecolor{codegray}{rgb}{0.5,0.5,0.5}
\definecolor{codepurple}{rgb}{0.58,0,0.82}
\definecolor{backcolour}{rgb}{0.95,0.95,0.92}

\lstdefinestyle{mystyle}{
    backgroundcolor=\color{backcolour},   
    commentstyle=\color{codegreen},
    keywordstyle=\color{magenta},
    numberstyle=\tiny\color{codegray},
    stringstyle=\color{codepurple},
    basicstyle=\ttfamily\footnotesize,
    breakatwhitespace=false,         
    breaklines=true,                 
    captionpos=b,                    
    keepspaces=true,                 
    numbers=left,                    
    numbersep=5pt,                  
    showspaces=false,                
    showstringspaces=false,
    showtabs=false,                  
    tabsize=2
}
\newcolumntype{P}[1]{>{\centering\arraybackslash}p{#1}}  %% Se crea un nuevo tipo de columna llamada P.

\graphicspath{ {Figs/} }  %%Ruta donde se encuentran las imágenes, que esté vacio indica que las imagenes están dentro de la misma carpeta que contiene el archivo .tex

\begin{document}

%%%%%%%%%%%%%%%%%%%%%%%%%%%%
%%% TÍTULO DEL DOCUMENTO %%%
%%%%%%%%%%%%%%%%%%%%%%%%%%%%
\title{\vspace{0.4cm}\Large Computation of 3D Band Structure and Density of States (DOS) of 14 Face Centered Cubic (FCC) Crystals using Pseudopotentials}
\author{Mirza Akbar Ali \\ Mirza-Akbar\_Ali@etu.u-bourgogne.fr \\ Quantum Technolologies and Engineering Erasmus Mundus Master (quanTEEM) \\ Université de Bourgogne, Dijon, France }

%%%%%%%%%%%%%%%%%%%%%%%%%%%%
%%%%%%%%% AUTORES %%%%%%%%%
%%%%%%%%%%%%%%%%%%%%%%%%%%%

%%%%%%%%%%%%%%%%%%%%%%%%%%%

% Comando que indica la generación del título
\maketitle
\thispagestyle{fancy}
%%%%%%%%%%%%%%%%%%%%%
%%%%%% RESUMEN %%%%%%
%%%%%%%%%%%%%%%%%%%%%
%\begin{abstract}
%En este documento se presenta una plantilla elaborada en \LaTeX para la presentación del documento final, como producto del desarrollo de la asignatura práctica de ingeniería electrónica. En esta sección se espera el resumen del problema, la solución propuesta y los principales resultados alcanzados.
%\end{abstract}
% En el resumen no se recomienda colocar citaciones bibliográficas.
%%%%%%%%%%%%%%%%%%%%%%
%%% PALABRAS CLAVE %%%
%%%%%%%%%%%%%%%%%%%%%%
%%%%%%%%%%%%%%%%%%%%%%
%\IEEEpeerreviewmaketitle
%%%%%%%%%%%%%%%%%%%%%%%%%%%%%%%%%%%%%
%%% PRIMERA SECCIÓN DEL DOCUMENTO %%%
%%%%%%%%%%%%%%%%%%%%%%%%%%%%%%%%%%%%%
\section{Introduction}
\IEEEPARstart{e}{lectronic} properties of materials are crucial to their ability to function in a wide range of applications, from electronics and energy production to structural materials and biomedicine\cite{HASEEB2016}. In order to understand and predict these properties, it is essential to study the energy levels and electron distribution within the material, which can be done through the study of the band structure and density of states (DOS)\cite{ashcroft1976solid}.

\par The band structure of a material describes the allowed energy levels that electrons can occupy within the material. These energy levels are formed by the interaction of the electrons with the crystal lattice and other electrons in the material. They are typically organized into bands separated by energy gaps. The band structure can be calculated using various theoretical methods, such as density functional theory (DFT) and the tight-binding model, and it can also be measured experimentally using techniques such as angle-resolved photoemission spectroscopy (ARPES)\cite{ashcroft1976solid}.

\par The density of states (DOS) is a measure of the number of electronic states available at a given energy level. It is typically calculated by considering the electronic states within a particular energy range, such as a band or an energy gap. The DOS can be used to calculate a range of electronic properties, such as the electrical conductivity and thermoelectric power, as well as to understand the electronic and optical responses of the material\cite{martin2004}.

\par Both the band structure and DOS are important for understanding the electronic properties of materials and predicting their behavior in various applications. For example, the band structure can be used to predict the electrical conductivity of a material, as well as its suitability for use in electronics and optoelectronics. The DOS can be used to understand the optical properties of a material, such as its absorption coefficient and refractive index, as well as to predict its thermoelectric properties.

\par In addition to their theoretical and predictive capabilities, the study of the band structure and DOS also has important practical applications. For example, the band structure can design new materials with specific electronic properties, such as high electrical conductivity or optoelectronic performance. The DOS can be used to optimize materials for energy conversion and storage applications, such as thermoelectrics and batteries\cite{Jaros1991}.

\par Overall, the study of the band structure and DOS is an essential part of materials science and engineering, with wide-ranging implications for various fields and applications. In this article, we will compute the three-dimensional electronic band structure and density of states (DOS) of 14 face-centred cubic (FCC) crystals using pseudopotentials.

%%%%%%%%%%%%%%%%%%%%%%%%%%%%%%%%%%%%%
%%%%% SECCIONES DE MARCO TEÓRICO %%%%
%%%%%%%%%%%%%%%%%%%%%%%%%%%%%%%%%%%%%
\section{Methodology}	
\par We adopt Bloch Model for the calculations of the band structures. The Bloch Model describes the spectrum of the electron energy states in the framework of the one‐electron approximation using a periodic potential that is independent of time. 
\subsection{Bloch Model}\label{BlochModelSec}
\par Bloch's theorem applies to wave functions of electrons inside a crystal and rests in the fact that the Coulomb potential in a crystalline solid is periodic. As a consequence, the potential energy function, $V(\Vec{r})$, in Schrödinger's equation should be of the form:
\begin{align}
    V(\Vec{r}) = V(\Vec{r} + \Vec{R}_n) \label{PeriodicPotential}
\end{align}
Here, $n$ points to a triplet of integer numbers $(n_1, n_2, n_3)$ identifying a vector of the direct lattice $\vec{R}_n$, which is itself expanded on the direct lattice basis vectors $\Vec{a}_1, \Vec{a}_2, \Vec{a}_3$ according to:
\begin{align}
    \Vec{R}_n = n_1\Vec{a}_1 + n_2\Vec{a}_2 + n_3\Vec{a}_3
\end{align}
The periodic $V(\Vec{r})$ given in equation \ref{PeriodicPotential} may be expanded by a FOURIER series using the reciprocal lattice vector $\vec{G}$:
\begin{align}
    V(\vec{r}) = \sum_{\vec{G}}V_{\vec{G}}e^{i\vec{G}\cdot\vec{r}}
\end{align}
with the form factor of FCC lattice defined by
\begin{align}
    V_{\vec{G}} = V_{|\vec{G}|^2} ^{S} \cos (\vec{G}\cdot\vec{\tau}) + iV_{|\vec{G}|^2} ^{S} \cos (\vec{G}\cdot\vec{\tau}) \label{V_G_Def}
\end{align}
Where, with lattice spacing $a$, $\vec{\tau} = \frac{a}{8}(1,1,1)$ and $-\vec{\tau}$ are the positions of the two atoms relative to the centre of the primitive cell that is chosen as the origin.
\par The values of $V_{|\vec{G}|^2} ^{S}$ and $V_{|\vec{G}|^2} ^{A}$ are deduced from fit to experimental data. They feature $V_{\vec{G}}=0$ if $|\vec{G}|^2 > 11 (4\pi^2/a^2 \,\text{units})$.For a selection of 14 semiconductor, Table \ref{FormFactors} shows the non-zero values for $|\vec{G}|^2\leq 11 (4\pi^2/a^2\text{ units})$. Even if the Fourier expansion of the potential is limited to $|\vec{G}|^2\leq 11 (4\pi^2/a^2 \,\text{units})$, a satisfactory converges requires that the representation of the Schr\"odinger equation in reciprocal space involves all $\vec{G}$ vectors such that $|\vec{G}|^2\leq 21 (4\pi^2/a^2 \,\text{units})$.
\begin{table}
    \caption{Lattice constant $a$ in angstroms and Pseudopotential form factors, in rydberg, derived from the experimental energy band splittings\cite{PhysRev.141.789}.}
    \label{FormFactors}
    \centering
    \begin{tabular*}{\linewidth}{l@{\extracolsep{\fill}}ccccccc} \hline\hline \\ \vspace{0.1cm}
    &$a$& $V_3^S$ & $V_8^S$ & $V_{11}^S$ & $V_3^A$ & $V_4^A$ & $V_{11}^A$ \vspace{0.1cm} \\ \hline\hline\\
    Si &5.43& -0.21 & 0.04 & 0.08 & 0.00 & 0.00 & 0.00 \\
    Ge &5.66& -0.23 & 0.01 & 0.06 & 0.00 & 0.00 & 0.00 \\
    Sn &6.49& -0.20 & 0.00 & 0.04&  0.00 & 0.00 & 0.00 \\
    GaP &5.44& -0.22 & 0.03 & 0.07 & 0.12 & 0.07 & 0.02 \\
    GaAs &5.64& -0.23 & 0.01 & 0.06 & 0.07 & 0.05 & 0.01 \\
    AlSb &6.13& -0.21 & 0.02 & 0.06 & 0.06 & 0.04 & 0.02 \\
    InP &5.86& -0.23 & 0.01 & 0.06 & 0.07 & 0.05 & 0.01 \\
    GaSb &6.12& -0.22 & 0.00 & 0.05& 0.06& 0.05& 0.01\\
    InAs &6.04& -0.22& 0.00& 0.05& 0.08& 0.05& 0.03 \\
    InSb &6.48& -0.20& 0.00& 0.04& 0.06& 0.05& 0.01\\
    ZnS &5.41&  -0.22& 0.03& 0.07& 0.24& 0.14& 0.04\\
    ZnSe &5.65& -0.23& 0.01& 0.06& 0.18& 0.12& 0.03\\
    ZnTe &6.07&  -0.22& 0.00& 0.05& 0.13& 0.10& 0.01\\
    CdTe &6.41& -0.20& 0.00& 0.04& 0.15& 0.09& 0.04\\ \hline\hline
    \end{tabular*}
\end{table}
\par The potential $V(\vec{r})$ discussed above, being independent of time, allows for the separation of variables of the Schr\"odinger equation. Thus, finding the eigenenergies consists in solving the time-independent Schr\"odinger equation for wave function $\psi(\vec{r})$ expanded on the basis of Sommerfield free electron model given by:
\begin{align}
    \psi(\vec{r}) = \frac{1}{\mathcal{V}}\sum_{\vec{k}}c_{\vec{k}}e^{i\vec{k}\cdot\vec{r}}
\end{align}
We find the expression of Schr\"odinger equation in reciprocal space:
\begin{align}
    \left( \frac{\hbar^2q^2}{2m_e}-E\right)c_{\vec{q}}+\sum_{\vec{G}}V_{\vec{G}}c_{\vec{q}-\vec{G}} = 0 \label{SE_Rec}
\end{align}
The system of equations given in equation \ref{SE_Rec} couples the values $\vec{q}$ differing from each other by a reciprocal lattice vector. Which, for $\vec{q} = \vec{k}-\vec{G}'$ and $\vec{G}'+\vec{G}=\vec{G}''$ can be written as 
\begin{align}
    [\bar{H}(\vec{k})-\mathbb{I}E]\bar{C}(\vec{k}) = 0
\end{align}
with $\bar{C}(\vec{k}) = 0$, and $\bar{H}(\vec{k})$ is the Hamiltonian matrix in reciprocal space given by:
\small{\begin{align}
    \left[ \begin{matrix}
        \ddots & \vdots & \vdots & \vdots & \vdots & \vdots& \iddots \\
        \cdots & T_{\vec{k}+\vec{G}_2} & V_{\vec{G}_1} & V_{\vec{G}_2} & V_{\vec{G}_3} & V_{\vec{G}_4} & \cdots \\
        \cdots & V_{\vec{G}_{-1}} & T_{\vec{k}+\vec{G}_1} & V_{\vec{G}_1} & V_{\vec{G}_2} & V_{\vec{G}_3} & \cdots \\
        \cdots & V_{\vec{G}_{-2}} & V_{\vec{G}_{-1}}  & T_{\vec{k}-0} & V_{\vec{G}_1} & V_{\vec{G}_2}  & \cdots\\
        \cdots & V_{\vec{G}_{-3}} & V_{\vec{G}_{-2}} & V_{\vec{G}_{-1}}  & T_{\vec{k}-\vec{G}_1} & V_{\vec{G}_1}  & \cdots\\
        \cdots & V_{\vec{G}_{-4}}  & V_{\vec{G}_{-3}} & V_{\vec{G}_{-2}} & V_{\vec{G}_{-2}}  & T_{\vec{k}-\vec{G}_1} & \cdots\\
        \iddots & \vdots & \vdots & \vdots & \vdots & \vdots& \ddots
    \end{matrix}
    \right]
\end{align}}
with \begin{align*}
    T_{\vec{k}-\vec{G}'} = \frac{\hbar^2 |\vec{k}-\vec{G}'|^2}{2m_e} + V_{\vec{G}=0}
\end{align*}
\par Since $V(\vec(r)$ are real, it allows us to deduce $V_{-\vec{G}} =V_{-\vec{G}}^{*}$, which implies that the hamiltonian matrix is Hermitian. This coupled Hamiltonian matrix can be decoupled by setting $V_{\vec{G}=0} = V_0 = 0$, which corresponds to the constant adjusting the zero of the potential\cite{SolidNotes}. Thus plotting the band structure involves finding and plotting the eigenvalues of the decoupled Hamiltonian matrix.
\subsection{Program Implementation for Band Structure Calculations}\label{ProgramImplementation}
\par Complete program for the calculation of the band structures can be divided into the following steps.
\begin{description}
\item[Step 1: ] Define the number of bands to be plotted, cutoff and maximum value of $\vec{G}$ described in \ref{BlochModelSec}, lattice spacing and Pseudopotential Form Factors provided in Table \ref{FormFactors}, positions of atoms $(\tau\text{ and } -\tau)$ in the primitive cell.
\item[Step 2: ] Generate FCC lattice unit vectors in cartesian coordinates and volume of the primitive cell.
\item[Step 3: ] Generate FCC reciprocal lattice unit vectors in cartesian coordinates. Calculate the minimum norm of the  reciprocal lattice unit vectors and defines the number of positive steps along each reciprocal lattice unit vector. Define the number of the reciprocal lattice vectors to be generated.
\item[Step 4: ] Generating reciprocal lattice vectors for calculations in all directions, sorting reciprocal lattice vectors by growing norm and keeping the reciprocal lattice vectors less than the cutoff limit.
\item[Step 5: ] Generate Brillouin Zone (BZ) exploration path according to traditional solid-state representation. 
\item[Step 6: ] For the kept reciprocal lattice vectors, calculate the value of $V_\vec{G}$ defined in Eq \ref{V_G_Def}.
\item[Step 7: ] Initialize Hamiltonian matrix and assign potential energy values corresponding to $\vec{G}$.
\item[Step 8: ] Calculate the difference $|\vec{k}-\vec{G}|^2$ along the BZ exploration path and kinetic energy part of the hamiltonian matrix, then diagonalize to find the eigenenergies.
\item[Step 9: ] Plot the eigenenergies against the BZ exploration path to obtain band structure. 
\end{description}
\subsection{Density of States (DOS)}
\par Density of states refers to the number of states available per unit of energy, thus having units $[eV^{-1}]$. Unlike band structure calculations, DOS calculation is not restricted to exploring just the high symmetry path in BZ. We define a sequence of numbers
\begin{align}
    u_r = (2r-q-1)/2q \quad \quad (r=1,2,3,\cdots,q)
\end{align}
Where $q$ is an integer that determines the number of special points in the set. With the above $u_r$'s we define 
\begin{align}
    \vec{k}_{prs} = u_p\vec{b}_1 + u_r\vec{b}_2 + u_s\vec{b}_3 \label{MonkhorstPoints}
\end{align}
That gives $q^3$ distinct points in reciprocal space uniformly distributed in BZ\cite{Monkhorst}. Then, points in rest of the octants are generated by symmetry. We follow the same steps as outlined in \ref{ProgramImplementation}, except for the Step 5. Instead of exploring the BZ path along high symmetry points, we now perform the calculations for the points generated by Eq \ref{MonkhorstPoints}. Once eigenenergies have been calculated for all the points in BZ, we perform calculations for the DOS, which is to be understood as a distribution \cite{SolidNotes}, given by
\begin{align}
    \mathcal{D}(E) = \sum_{\vec{k}}\delta(E-E_{\vec{k}}) \label{DOSDistribution}
\end{align}
where 
\begin{align}
    \delta(E-E_{\vec{k}})= \frac{e^{-(E-E_{\vec{k}})^2/\sigma^2}}{\sigma\sqrt{\pi}} 
\end{align} 
\section{Results}
\par Band structure and density of states (DOS) plots for energy range $-14\,eV$ to $6\,eV$ for 14 face centered cubic (FCC) crystals produced with the developed program are shown in Figures 1-28.
\bibliographystyle{unsrt}
\bibliography{references.bib}

\begin{thebibliography}{1}

\bibitem{HASEEB2016}
A.S.M.A. Haseeb.
\newblock Electronic materials.
\newblock In {\em Reference Module in Materials Science and Materials
  Engineering}. Elsevier, 2016.

\bibitem{ashcroft1976solid}
Neil~W Ashcroft and N~David Mermin.
\newblock Solid state physics, cornell university, 1976.

\bibitem{martin2004}
Richard~M. Martin.
\newblock {\em Electronic Structure: Basic Theory and Practical Methods}.
\newblock Cambridge University Press, 2004.

\bibitem{Jaros1991}
M.~Jaros.
\newblock {\em Concepts and Applications of Band Structure Engineering in
  Optoelectronics}, pages 147--163.
\newblock Springer US, Boston, MA, 1991.

\bibitem{PhysRev.141.789}
Marvin~L. Cohen and T.~K. Bergstresser.
\newblock Band structures and pseudopotential form factors for fourteen
  semiconductors of the diamond and zinc-blende structures.
\newblock {\em Phys. Rev.}, 141:789--796, Jan 1966.

\bibitem{SolidNotes}
Alain Dereux.
\newblock {\em Selected Chapters of Solid State Physics}.
\newblock Université de Bourgogne, 2022.

\bibitem{Monkhorst}
Hendrik~J. Monkhorst and James~D. Pack.
\newblock Special points for brillouin-zone integrations.
\newblock {\em Phys. Rev. B}, 13:5188--5192, Jun 1976.

\end{thebibliography}
\begin{figure}[p]
    \centering
    \include{Graphics/Si-3D-EK-Diagram.tex}
    \label{SiBS}
\end{figure}
\begin{figure}[p]
    \centering
    \include{Graphics/Ge-3D-EK-Diagram.tex}
    \label{GeBS}
\end{figure}
\begin{figure}[p]
    \centering
    \include{Graphics/Sn-3D-EK-Diagram.tex}
    \label{SnBS}
\end{figure}

\begin{figure}[p]
    \centering
    \include{Graphics1/E1-Si-mq80-smear0.05-DOSdata.tex}
    \label{DOSSi}
\end{figure}
\begin{figure}[p]
    \centering
    \include{Graphics1/E2-Ge-mq80-smear0.05-DOSdata.tex}
    \label{DOSGe}
\end{figure}
\begin{figure}[p]
    \centering
    \include{Graphics1/E3-Sn-mq80-smear0.05-DOSdata.tex}
    \label{DOSSn}
\end{figure}

\begin{figure}[p]
    \centering
    \include{Graphics/GaP-3D-EK-Diagram.tex}
    \label{GaPBS}
\end{figure}
\begin{figure}[p]
    \centering
    \include{Graphics/GaAs-3D-EK-Diagram.tex}
    \label{GaAsBS}
\end{figure}
\begin{figure}[p]
    \centering
    \include{Graphics/AlSb-3D-EK-Diagram.tex}
    \label{AlSbBS}
\end{figure}

\begin{figure}[p]
    \centering
    \include{Graphics1/E4-GaP-mq80-smear0.05-DOSdata.tex}
    \label{DOSGaP}
\end{figure}
\begin{figure}[p]
    \centering
    \include{Graphics1/E5-GaAs-mq80-smear0.05-DOSdata.tex}
    \label{DOSGaAs}
\end{figure}
\begin{figure}[p]
    \centering
    \include{Graphics1/E6-AlSb-mq80-smear0.05-DOSdata.tex}
    \label{DOSAlSb}
\end{figure}

\begin{figure}[p]
    \centering
    \include{Graphics/E7InP-3D-EK-Diagram.tex}
    \label{InPBS}
\end{figure}
\begin{figure}[p]
    \centering
    \include{Graphics/E8GaSb-3D-EK-Diagram.tex}
    \label{GaSbBS}
\end{figure}
\begin{figure}[p]
    \centering
    \include{Graphics/E9InAs-3D-EK-Diagram.tex}
    \label{InAsBS}
\end{figure}

\begin{figure}[p]
    \centering
    \include{Graphics1/E7-InP-mq80-smear0.05-DOSdata.tex}
    \label{DOSInP}
\end{figure}
\begin{figure}[p]
    \centering
    \include{Graphics1/E8-GaSb-mq80-smear0.05-DOSdata.tex}
    \label{DOSGaSb}
\end{figure}
\begin{figure}[p]
    \centering
    \include{Graphics1/E9-InAs-mq80-smear0.05-DOSdata.tex}
    \label{DOSInAs}
\end{figure}

\begin{figure}[p]
    \centering
    \include{Graphics/E10InSb-3D-EK-Diagram.tex}
    \label{InSbBS}
\end{figure}
\begin{figure}[p]
    \centering
    \include{Graphics/E11ZnS-3D-EK-Diagram.tex}
    \label{ZnSBS}
\end{figure}
\begin{figure}[p]
    \centering
    \include{Graphics/E12ZnSe-3D-EK-Diagram.tex}
    \label{ZnSeBS}
\end{figure}

\begin{figure}[p]
    \centering
    \include{Graphics1/E10-InSb-mq80-smear0.05-DOSdata.tex}
    \label{DOSInSb}
\end{figure}
\begin{figure}[p]
    \centering
    \include{Graphics1/E11-ZnS-mq80-smear0.05-DOSdata.tex}
    \label{DOSZnS}
\end{figure}
\begin{figure}[p]
    \centering
    \include{Graphics1/E12-ZnSe-mq80-smear0.05-DOSdata.tex}
    \label{DOSZnSe}
\end{figure}

\begin{figure}[p]
    \centering
    \include{Graphics/E13ZnTe-3D-EK-Diagram.tex}
    \label{ZnTeBS}
\end{figure}
\begin{figure}[p]
    \centering
    \include{Graphics/E14CdTe-3D-EK-Diagram.tex}
    \label{CdSeBS}
\end{figure}
\begin{figure}[p]
    \centering
    \include{Graphics1/E13-ZnTe-mq80-smear0.05-DOSdata.tex}
    \label{DOSZnTe}
\end{figure}
\begin{figure}[p]
    \centering
    \include{Graphics1/E14-CdTe-mq80-smear0.05-DOSdata.tex}
    \label{DOSCdTe}
\end{figure}
\newpage \onecolumn
\section{Band structures showing all 16 Bands for 14 materials}
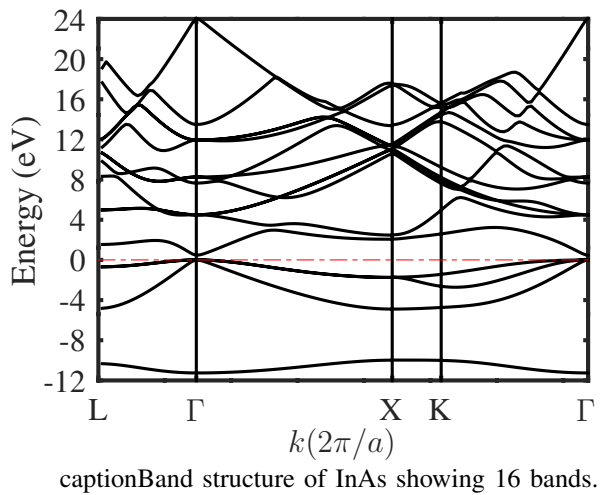
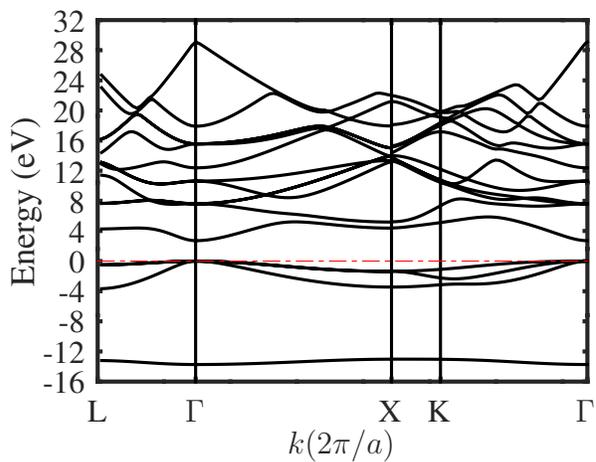
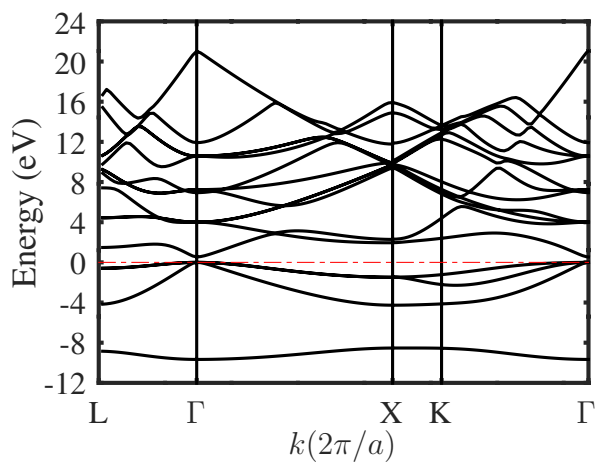
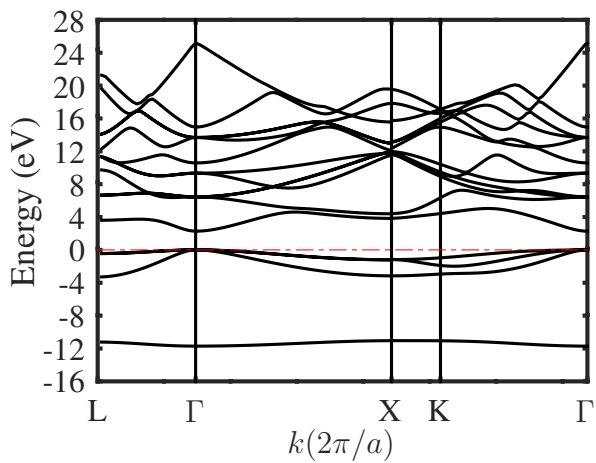
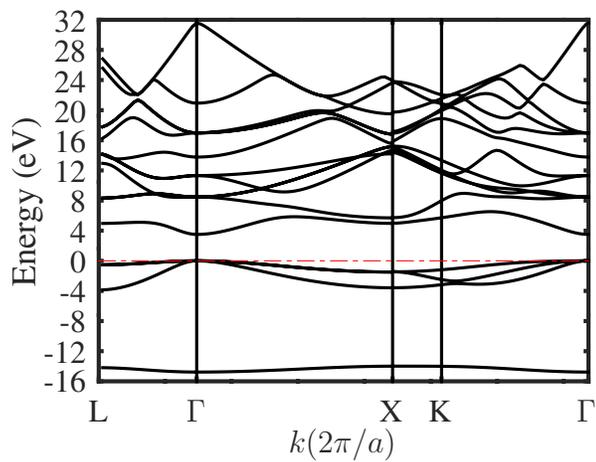
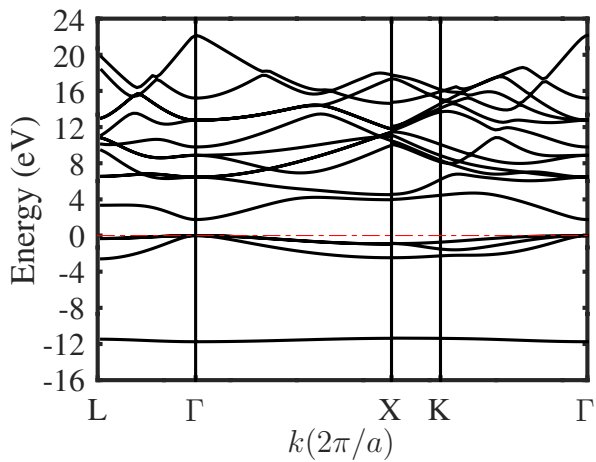
\vfill
\begin{figure}[h]
    \centering
    \include{Graphics3/E1Si-3D-EK-Diagram.tex}
    \label{SiBSComplete}
\end{figure}
\vfill
\begin{figure}[h]
    \centering
    \include{Graphics3/E2Ge-3D-EK-Diagram.tex}
    \label{GeBSComplete}
\end{figure}
\vfill
\twocolumn
\begin{figure}[p]
    \centering
    \include{Graphics3/E3Sn-3D-EK-Diagram.tex}
    \label{SnBSComplete}
\end{figure}

\begin{figure}[p]
    \centering
    \include{Graphics3/E4GaP-3D-EK-Diagram.tex}
    \label{GaPBSComplete}
\end{figure}
\begin{figure}[p]
    \centering
    \include{Graphics3/E5GaAs-3D-EK-Diagram.tex}
    \label{GaAsBSComplete}
\end{figure}
\begin{figure}[p]
    \centering
    \include{Graphics3/E6AlSb-3D-EK-Diagram.tex}
    \label{AlSbBSComplete}
\end{figure}

\begin{figure}[p]
    \centering
    \include{Graphics3/E7InP-3D-EK-Diagram.tex}
    \label{InPBSComplete}
\end{figure}
\begin{figure}[p]
    \centering
    \include{Graphics3/E8GaSb-3D-EK-Diagram.tex}
    \label{GaSbBSComplete}
\end{figure}
\begin{figure}[p]
    \centering
    \include{Graphics3/E9InAs-3D-EK-Diagram.tex}
    \label{InAsBSComplete}
\end{figure}

\begin{figure}[p]
    \centering
    \include{Graphics3/E10InSb-3D-EK-Diagram.tex}
    \label{InSbBSComplete}
\end{figure}
\begin{figure}[p]
    \centering
    \include{Graphics3/E11ZnS-3D-EK-Diagram.tex}
    \label{ZnSBSComplete}
\end{figure}
\begin{figure}[p]
    \centering
    \include{Graphics3/E12ZnSe-3D-EK-Diagram.tex}
    \label{ZnSeBSComplete}
\end{figure}
\begin{figure}[p]
    \centering
    \include{Graphics3/E13ZnTe-3D-EK-Diagram.tex}
    \label{ZnTeBSComplete}
\end{figure}
\begin{figure}[p]
    \centering
    \include{Graphics3/E14CdTe-3D-EK-Diagram.tex}
    \label{CdSeBSComplete}
\end{figure}
\newpage
\onecolumn
%\section{Octave Scripts}
%\subsection{Start Defaults (mystartdefaults.m)}
%\lstinputlisting{mystartdefaults.m}
%\subsection{Brillouin Zone (BZ) Exploration Path}
%\par \noindent Generate Brillouin Zone (BZ) exploration path according to traditional solid-state representation for FCC. bzpath.dat file generated by this code is used in band structure calculations.
%\lstinputlisting{BZ_FCC.m}
%\subsection{Code for Band Structure Calculations}
%\lstinputlisting{BandStructureMain.m}
%\subsection{Code for Band Structure Plotting (BS\_Plot.m)}
%\lstinputlisting{BS_Plot.m}
%\subsection{Code for Density of States (DOS) Calculations}
%\lstinputlisting{DOS_Main.m} \newpage
%\subsection{Code for Density of States (DOS) Plotting}
%\lstinputlisting{DOS_Plot.m}
%%%%%%%%%%%%%%%%%%%%%%%%%%%%%%%%%%%%%
\newpage
\onecolumn
\ifCLASSOPTIONcaptionsoff
  \newpage
\fi

%%%%%%%%%%%%%%%%%%%%%%%%%%
%%%%%% BIBLIOGRAFIA %%%%%%
%%%%%%%%%%%%%%%%%%%%%%%%%%

%%%%%%%%%%%%%%%%%%%%%%%%%%
\newpage
\centering{\textit{\Large{Please contact the document's author for the availability of the band structures in High Resolution. Figures can be used for academic purposes with permission.}}}
\end{document}